\documentstyle{l-aa}
\begin{document}
\thesaurus{}
\title{An Interferometric Study of the Blue Compact Dwarf Galaxy IZW18
\thanks{Based on observations performed with the Canada-France-Hawaii 
Telescope}}
\author{A. R. Petrosian 
\inst{1}
\and J. Boulesteix 
\inst{2}
\and G. Comte 
\inst{2}
\and D. Kunth
\inst{3}
\and E. LeCoarer
\inst{4}}

\offprints{D. Kunth}

\institute{Byurakan Astrophysical Observatory, Armenia
\and Observatoire de Marseille, 2 Place le Verrier, 
 F-13248 Marseille Cedex 04, France
\and Institut d'Astrophysique de Paris, 98bis. Boulevard Arago, 75014, Paris,
France
\and Observatoire de Grenoble, Universite Joseph Fourier, 
414, rue de la Piscine, Boite Postale 53X, F-38041, Grenoble Cedex, France} 

\date{Received ...........1996; accepted .............1996}

\maketitle 

\begin{abstract} 
We present high spatial resolution observations of the blue compact
dwarf galaxy IZW18 performed in the  H$\alpha$ line with a scanning  Fabry-Perot
interferometer at the CFH telescope. Morphological structure of
the galaxy in H$\alpha$  and in the red continuum is investigated.
We also analyse the
velocity field of the ionized gas.
Besides the two compact HII components of the main body we find a population
 of small HII regions  in its surroundings
whose diameter distribution and H$\alpha$ luminosity function are 
consistent with those observed in dwarf irregular galaxies. In the main 
body of the galaxy besides of the NW and SE red continuum peaks which are 
displaced with respect to the H$\alpha$ maxima, three new red condensations 
have been discovered. They have no clear H$\alpha$ counterparts. The 
velocity field in IZW18 shows peculiar motions superimposed on a quite regular 
background implying solid-body rotation with a gradient of about 
70~km~$\rm s^{-1}$ kpc$\rm ^{-1}$. The H$\alpha$ line profiles 
exhibit an asymmetric structure, except for the NW main compact component. 
At least part of this asymmetry could result from accreted and/or expelled
surrounding gas from the main star-forming core(s) of the
galaxy. Contrary
 to previous suggestions that the south-west
and north-east extensions of this galaxy are diffused emission produced by
 bipolar emitting gas we provide evidence that they are HII regions powered by
star formation sites. The redshift of the  Zwicky's "flare"
 has been measured for the first time and corresponds to the same velocity as
 IZW18. In such a context the  optical ridge that appears to be an
isolated
 morphological structure has a shape that may result from the gravitational
 interaction with the  Zwicky's "flare" if this latter is a neighbour extreme
dwarf object.

\end{abstract}

\section{Introduction}

Extensive studies of Blue Compact Dwarf Galaxies (BCDGs) have
revealed that a large fraction of them are chemically unevolved
small systems undergoing violent star formation activity (Thuan
1987). The triggering mechanism of starbursts in BCDGs is still
completely unknown. One of the approaches for understanding this
mechanism is to investigate the morphological,
kinematical and dynamical relationship between star--forming regions 
and their surrounding gas and also to search for global correlations
between star--forming regions and the gas in various places of a
galaxy. Since BCDGs are among the best candidates for genuine
zero-redshift  young galaxies their kinematical and dynamical
state may give also some clues on the nature of dark matter
(Tremaine \& Gunn 1979).

         The neutral gas component of some  BCDGs has been kinematically
studied with radio interferometers (Viallefond \& Thuan 1983;
Viallefond et al. 1987; Taylor et al. 1991). In the few objects studied so
far, this gas shows core-halo
 distribution
 and some chaotic velocity superimposed on a regular velocity field
 indicative of a solid body rotation.

        On the other hand the ionized gas component in BCDGs has a clumpy
 distribution and is mainly concentrated in at least one compact star--forming
HII region immersed in an amorphous extended envelope of
low luminosity ionized  material (Kunth et al. 1988). The inner
dynamics of HII regions in BCDGs has been investigated by
Gallagher \& Hunter (1983) and Lequeux et al. (1995). Velocity 
fields in BCDGs have been mapped using multi-slit
observations (Tomita et al. 1993) and H$\alpha$ Fabry-Perot
interferometric observations (Thuan et al. 1987).

         The combination of a Fabry-Perot interferometer with a large 
telescope provides high spatial and velocity resolution and allows to 
probe small scales in dwarf compact galaxies, yielding information on
components not resolved on HI maps (Thuan et al. 1987). Further, a BCDG not 
only has a small angular size and a compact distribution  but at the same time
exhibits very strong and narrow 
emission lines. For this reason BCDGs are ideal targets for  Fabry-Perot
interferometer observations. 
In this paper we present results relevant to the extreme
metal--poor BCDG IZW18 (Mrk116).

         IZW18 first was described by Zwicky (1966) as a "double
system of compact galaxies". Later IZW18 became an outstanding
extragalactic object when Sargent \& Searle (1970) characterized
it as "isolated extragalactic HII region" and found that it is
extremely metal poor as compared to the Sun (Searle \& Sargent 1972). 
Later on after 25 years more than 100 scientific papers have been addressed 
to this galaxy. But among them, articles which contain data on IZW18 as a 
extended two component system (Petrosian et al. 1978; Mazzarella \& 
Boroson 1993) are rare. Optical imagery of IZW18 has been 
published by Hua et al. (1987), Davidson et al. (1989, hereafter DKF89), 
Dufour \& Hester (1990, hereafter DH90) and Martin (1996, hereafter M96).
Broadband WFPC2 on HST observations were carried out by Hunter \& Thronson 
(1995, hereafter HT95).  GHRS HST spectroscopic observations revealed very 
low oxygen abundance in the HI envelope of this galaxy (Kunth et al. 1994, 
but see Pettini \& Lipman 1995 for a critical comment of this result).  
HI line interferometric observations by Lequeux \& Viallefond (1980), 
Viallefond et al. (1987) and radio continuum observations by Klein et 
al. (1983) were also reported. Probably due to its very low metal-content
IZW18 has been detected neither by IRAS 
(Kunth \& Sevre 1986; Mazzarella et al. 1991) nor in CO millimeter line 
(Arnault et al. 1988).  

         As part of our high spatial and spectral resolution 
study of genuine BCDGs, in this paper we present  the H$\alpha$ scanning 
Fabry-Perot interferometric observations of IZW18. We describe the 
observations and data reduction in Section 2. Results are presented  in 
Section 3 and are discussed in Section 4. We summarize our conclusions 
in Section 5. We shall adopt 10 Mpc as the distance to IZW18, based on a 
recessional velocity of 763~km~$\rm s^{-1}$ and a Hubble constant 
H$\rm _{o}$ = 75~km~$\rm s^{-1}$ Mpc$\rm ^{-1}$. The linear scale is 
therefore 48 pc arcsec$\rm ^{-1}$.

\section{Observations and data reduction}

         Observations have been performed at the Cassegrain focus of
the 3.6 m Canada-France-Hawaii telescope on November 22 1986,
with the CIGALE instrument (Boulesteix et al. 1983). It  consists
of a focal reducer (f/8 - f/2), and a piezo-electrically
scanned  Fabry - Perot interferometer from Queensgate Instruments 
interfaced with a two dimensional photon counting camera. 
The optical setup of the f/2
focal reducer allows a field of view of 6.\arcmin2 x 6.\arcmin2. The original
photon addresses are computed on a 512 x 512 pixel grid on the
detector,  giving a scale of  0.\arcsec725 per pixel.
 
         The Fabry-Perot etalon operates with an interference order
of 798 at the rest wavelength of H$\alpha$ providing a free velocity
range of 375~km~$\rm s^{-1}$. The corresponding free spectral range was scanned through
21 scanning steps (each step corresponding to a given spacing of
the plates of the interferometer), the 21th step being equivalent
to the 1st one. Thus the velocity sampling step is 18.8~km~$\rm s^{-1}$ .
Each scanning step has been exposed 18 x 10s. The total exposure
time was 60 minutes (18x10x20s). A neon lamp emitting the
6598.9~\AA\  
line was used for wavelength calibration.

         The final product of the 20 elementary
interferograms processing is a 3 - dimensional data cube with 2 spatial and
one velocity coordinates. For each pixel with reasonably high S/N ratio
 the radial velocity, the  H$\alpha$ line
 profiles and the   H$\alpha$ and
continuum fluxes are derived from this cube. 

         The night sky OH emission line at 6577.2~\AA~has a wavelength
value close to that of the redshifted IZW18 H$\alpha$ emission. An interactive
routine was used to subtract the night sky line. The radial
velocity at each pixel was then extracted by a barycentric
technique (Laval et al. 1987). The barycenter of the H$\alpha$ line is
found with an accuracy much better than the interferometer  scanning--step 
 (18.8~km~$\rm s^{-1}$).

         From the data cube, H$\alpha$ line profiles are built for each
pixel. Gaussian curves were fitted to the observed profiles to
investigate their structure and derive velocity dispersions
due to macroscopic gas motions. The velocity dispersion of the
Gaussian profile which best fits the observed profile
is defined by 
            
  $$\beta^2 = \beta^2_{\rm obs}-\beta^2_{\rm i}-\beta^2_{\rm th}$$ 
 
where $\beta_{\rm th}$ is the thermal broadening
function and $\beta_{\rm i}$ the instrumental profile. According to the 
value T$\rm _{e}$ = 18000 K of the nebular gas electron temperature of 
IZW18 (Dufour et al. 1988) the thermal broadening $\beta_{\rm th}$ is 
estimated as 12.2~km~$\rm s^{-1}$ (Dopita 1972). The
instrumental profile determined by scanning the neon 6598.95~\AA\
line, was found to be well fitted by a Gaussian with velocity
width $\beta_{\rm i}$ = 14.4~km~$\rm s^{-1}$. The observed profile is
described with $\beta_{\rm obs}$ = 0.425 FWHM.

         The total intensity profile for each pixel when
adding up the 21 channels, contains information about the
monochromatic emission as well as the continuum emission of the
object. CIGALE observations enable us to separate monochromatic
emission from continuum emission (see details in Laval et al.
1987). It is thus possible to restore pure monochromatic H$\alpha$ 
images of IZW18, completely free from the underlying stellar  continuum.
The
continuum subtracted H$\alpha$ line images of IZW18 were calibrated
using the absolute flux  published by Davidson \& Kinman (1985)
and DH90. In the same way a continuum map free from H$\alpha$ line
contribution was built. The FWHM of the interference filter was
10~\AA\ which determined the bandwidth of the image continuum of IZW18.
Continuum flux calibrations were based on the assumption that the
continuum slope of IZW18 is given by $$f(\lambda)  = A\lambda^{-2.5}$$ 
(Davidson \& Kinman
1985) and on absolute flux values at 6600~\AA\ 
given by DKF89. In making a continuum model we have neglected the
underlying stellar abosrption since according to  Skillman and Kennicutt (
1993) the absorption component in H$\alpha$ is of the order of 2\AA\ while the
H$\alpha$ emission has an equivalent width more than two order of magnitude
larger (465\AA\ ). For the confirmation of faint detected emission--line
 features such as small narrow-line
HII regions, a careful check was made on the series of channel maps: a feature
has been considered as certain ONLY if seen on at least
two consecutive channel maps. 

         H$\alpha$ and continuum luminosities are derived from the 
measured fluxes and are corrected for extinction using values 
adopted by Dufour et al. (1988). 

         Since the image of IZW18 is sufficiently small to be located in
the central part of the field of view where all the causes for sensitivity
variations  are reduced (optical system, filter combination, detector
 inhomogeneities) the original images of the
galaxy were not flat-fielded.

\section{Results}
  
         When the interferometer is scanning the object, the
monochromatic emission of this latter is modulated by the
interference pattern while the continuum emission remains
unaffected. This enables to distinguish line and continuum
emission with a much higher contrast than with simple images
obtained through interference filters.

         Figures 1 and 2 show the logarithmic contour maps of
respectively the red continuum emission and the monochromatic H$\alpha$ 
emission in absolute flux units across the main body of IZW18.
	The selectivity of the Fabry-P\'erot interferometer due to its high
finesse ( F=16 measured on calibration frames) allows a very clear separation
of the line and the continuum emissions. Furthermore, the overall system
(FP + focal reducer) benefits from a very low level of scattered light, much
below a percent, which ensures that the continuum map is free from
additional line emission.

\begin{figure*}
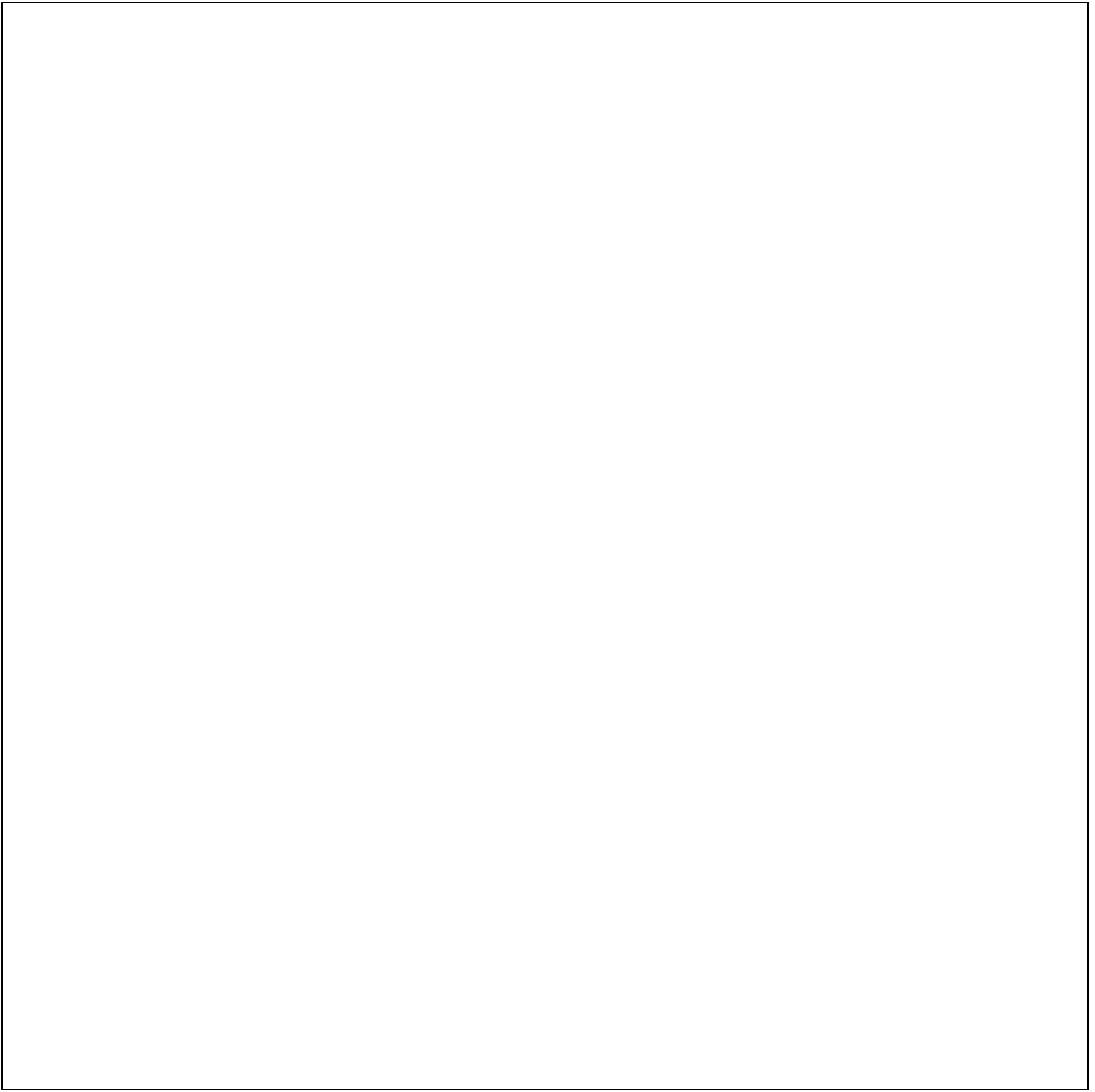

\picplace{18cm}
\caption[1]{The 6600\AA\ red continuum map of IZW18.
 The lowest contour corresponds to a flux of 1.2$10^{-18}$ergs~$\rm
 cm^{-2}~s^{-1}~A^{-1}$  per pixel. Contours are logarithmically spaced by
 a factor of 0.1. The insert is the detail view to central part
 of the galaxy. All morphological details are labelled.}
\end{figure*} 
 
\subsection{Quantitative red continuum morphology}

         In red continuum (Fig.1) the morphological structure 
of IZW18 is complex and new details appear which were not detected 
by DKF89. Some of these details have their counterparts on the HST WFPC2
frames  (HT95). In the main body of the galaxy besides  two known NW 
and SE red continuum components (DKF89) a third one (a) is observed half 
way between them close from the north of the HII region number 6
in HT95. We identify two other condensations: (b) to the north 
and (c) to the north-west of the NW compact red component. Among all the
components NW dominates in  size, surface and absolute brightness. Table 1 
lists $\Delta$$\alpha$, $\Delta$$\delta$ positions of the photometric center 
of the components in arcsec with respect to the NW component. Positive 
values of $\Delta$$\alpha$ are to the east and positive $\Delta$$\delta$ 
to the north, effective diameters D in parsecs, surface brightness SB in units of
10$^{-17}$ergs~$\rm cm^{-2}~s^{-1}~A^{-1}~arcsec^{-2}$  and red magnitudes 
at 6600~\AA. Table 1 also lists the same parameters for the  Zwicky's "flare" 
(Zwicky 1966) - noted as the C satellite  by DKF89 and DH90.

\begin{table*}
\caption[1]{Parameters for red continuum components}
\begin{flushleft}
\begin{tabular}{cccccc}
\hline\noalign{\smallskip} 
$Component$ & $\Delta \alpha$ & $\Delta \delta$ & $D$ & $SB(6600\AA)$ & $m(6600\AA)$ \\
   & \small{$\arcsec$} & \small{$\arcsec$} & \small{$pc$} & \small{$10^{-17} \ ergs \ cm^{-2}\ s^{-1}\ A^{-1}\ arcsec^{-2}$} \\
\hline\noalign{\smallskip}
$NW$ &  0   &  0   & 112 & 4.8 & 17.2 \\ 
$SE$ &  2.9 & -3.6 &  79 & 3.1 & 18.7 \\
$a$  &  1.5 & -1.5 &  56 & 3.2 & 19.2 \\
$b$  & -0.7 &  2.9 &  56 & 2.8 & 19.1 \\
$c$  & -2.9 &  1.5 &  40 & 4.0 & 20.2 \\
$C$  &-18.9 & 16.7 & 177 & 0.3 & 20.6 \\ 
\noalign{\smallskip}
\hline
\end{tabular}
\end{flushleft}
\end{table*}

         The comparison between the structure of the NW red component and its 
surroundings with that of HT95 frames shows that this component 
is coeval to an isolated HII region  south to 
the main NW star-forming complex (HT95). A sharp cut-off in the NW red
 continuum brightness 
distribution is observed towards the (a) and SE components  while it fades 
smoothly in the opposite direction towards the (b) component. The (b)
component itself is located in the northern part of NW star--forming 
complex of HT95. At faint levels of brightness a  west side "ledge" is 
observed. The (c) component is embedded into this ledge and  has no direct
counterpart on HST frames (HT95).

          Besides  the main body of the galaxy its outer envelope is also 
radiating in the red continuum. This radiation is not homogenous. It shows
a  clumpy structure  extending to the NE direction up to a 
 distance of about 740 pc 
from the NW component. Its patterns are located within the same region of the
NE H$\alpha$ shell of M96. Towards the SW the  red continuum radiation extends 
in more homogeneous way by curving at its end to the west and its 
patterns closely follow the H$\alpha$ arc--like structure  DH90.       

          Stellar light as well as some nebular continuum contribute to the
red continuum radiation of the outer envelope of the galaxy and its arc-like 
structure. To determine their respective  contribution  we  have measured the 
L(H$\alpha$)/L(6600\AA) ratios across these regions and 
compared the obtained values with  model 
calculations provided to us by Mas-Hesse (private communication). Models
show that 
 the underlying nebular continuum  should be less
than 1 percent of our  measured  luminosity at
6600 \AA\ .

\subsection{Quantitative H$\alpha$ morphology}

         Our H$\alpha$ monochromatic image (Fig.2) completely reproduces 
the structural features which were identified in earlier 
observations (Hua et al. 1987; DKF89; DH90; HT95; M96; Ostlin et 
al. 1996). Let us underline some details.

          We hereafter shall define an autonomous HII feature as the
region which is inside its  outermost closed contour. Both NW and SE 
condensations 
are isolated from the main body of the galaxy by an isophote corresponding 
to 4.5 10$^{-15}$ ergs~$\rm cm^{-2}~s^{-1}$ arcsec$\rm ^{-2}$ of surface 
brightness.

         The NW component of IZW18 has a  size of 195 x 117 pc$\rm ^{2}$,
while the SE
component is 157 x 117 pc$\rm ^{2}$. Note that the elongation of both
components coincides with the  elongation of the main body of the galaxy.

         After correction from Galactic reddening the integrated H$\alpha$ 
flux of 
NW and SE components come to be 1.3 10$^{-13}$ ergs~$\rm cm^{-2}~
s^{-1}$ and 6.3 10$^{-14}$ ergs~$\rm cm^{-2}~s^{-1}$ respectively. At 
D = 10 Mpc these values correspond to H$\alpha$ luminosities of 
1.6 10$^{39}$ergs~$\rm 
s^{-1}$ and 7.6 10$^{38}$ergs~$\rm s^{-1}$. Assuming Case B conditions and 
T$\rm _{e}$ = 18000 K (Dufour et al. 1988) these luminosities require 
conversion of 2.2 10$^{51}$ and 1.1 10$^{51}$  ionizing photons $\rm s^{-1}$\,
requiring 47 and 23 O5V stars, assuming that an O5V star emits 4.7 10$^{49}$ 
photons $\rm s^{-1}$\ (Osterbrock 1989).

         Star formation rates ( from 0.1 to 100$M_{\odot}$) in the NW 
and SE component  are equal  respectively to 4.7 10$^{-7}$$M_{\odot}$yr$\rm ^{-1}$ pc$\rm ^{-2}$ and 2.9 10$^{-7}$$M_{\odot}$yr$\rm ^{-1}$ pc$\rm ^{-2}$ 
assuming a Salpeter initial mass function (Hunter \& Gallagher 1986).

         Masses for the ionized gas in the NW and SE components have been
estimated to be M(HII) = 7.3 10$^{5}$$M_{\odot}$ for the NW condensation and 
M(HII) = 3.4 10$^{5}$$M_{\odot}$ for the SE one, but these standard  estimates
are only indicative since these HII regions are far from being homogeneous as
seen from the HST images.

\begin{figure*}
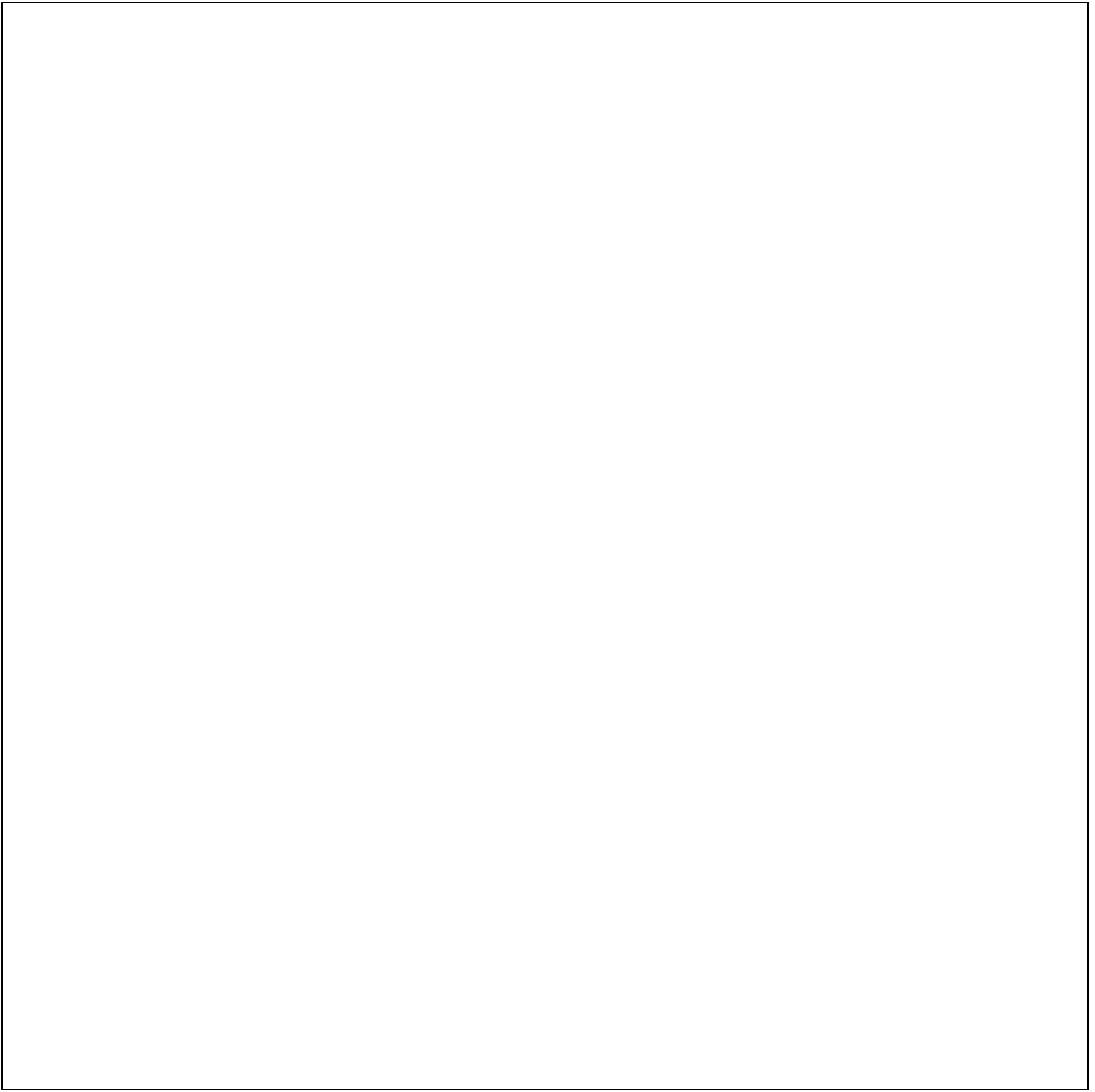

\picplace{18cm}
\caption[2]{H$\alpha$ intensity map of IZW18. The lowest contour corresponds 
to a flux of a $10^{-17}\ ergs~\rm cm^{-2} s^{-1}$ per pixel. Contours are 
logarithmically spaced by a factor of 0.2. HII regions and Zwicky's "flare" 
are labelled.}
\end{figure*} 
          
         The analysis of the mean spatial profiles of both components along
the direction of their elongation and perpendicular to it, shows
that the surface brightness peak  of the  NW H$\alpha$ component is about
7.0 10$^{-15}$ergs~$\rm cm^{-2}~s^{-1}$ arcsec$\rm ^{-2}$, as  compared to 
6.4 10$^{-15}$ergs~$\rm cm^{-2}~s^{-1}$ arcsec$\rm ^{-2}$ for the SE 
component. The H$\alpha$ surface brightness gradient on the NW component 
across two perpendicular directions is different. It is gently sloping across 
the major axis of the component and much steeper in the perpendicular 
direction where according to HT95 a gap in ionized gas distribution is 
observed. The smoothest  gradient is observed along the tail extending 
to the NW and wraps around to the west of the component. This feature 
was noted by DKF89, DH90 and also traced in the H$\alpha$ maps of Hua et 
al. (1987). It is very well seen in Fig.1b of M96 and shows its  
detailed structure in Fig. 3 of HT95. As regards to the SE compact component, 
its H$\alpha$ line brightness distribution is more homogeneous.

		 An interesting H$\alpha$ feature is seen as a  prominent 
winding ridge (DH90; HT95; M96; Ostlin et al. 1996) which expands south 
from the SE HII compact component, bends around to the NW , then turns 
back to the NE, making an half wrapping around the NW HII compact component.
 In the 
NW direction from  the north-western boundary of the ridge about the same 
distance as from the NW  component lies  Zwicky's "flare" (Zwicky
1966). The ridge as an 
isolated structure is well shaped and has a conspicuous  clumpy structure
(HT95). As mentioned above (Sec.3.1) several patterns of the ridge radiate
 some 
red continuum. The continuum patterns  do  not follow the  H$\alpha$ clumps 
distribution. To understand the nature of the ridge we carefully tested the 
possibility of its link to the SE compact component. We conclude that 
there is not direct connection between them. From this point of view the 
ridge can be an  isolated feature in the system. As it is noted by Viallefond 
et al. (1987) the two major star--forming areas in the main body of IZW18 are 
offset to the east of the double-lobed  HI emission peak. The ridge 
wraps around the HI distribution peak opposite to the side where 
the main body of the galaxy is located, roughly following the steep contours 
marking the edge of the HI concentration.   

\begin{figure*}
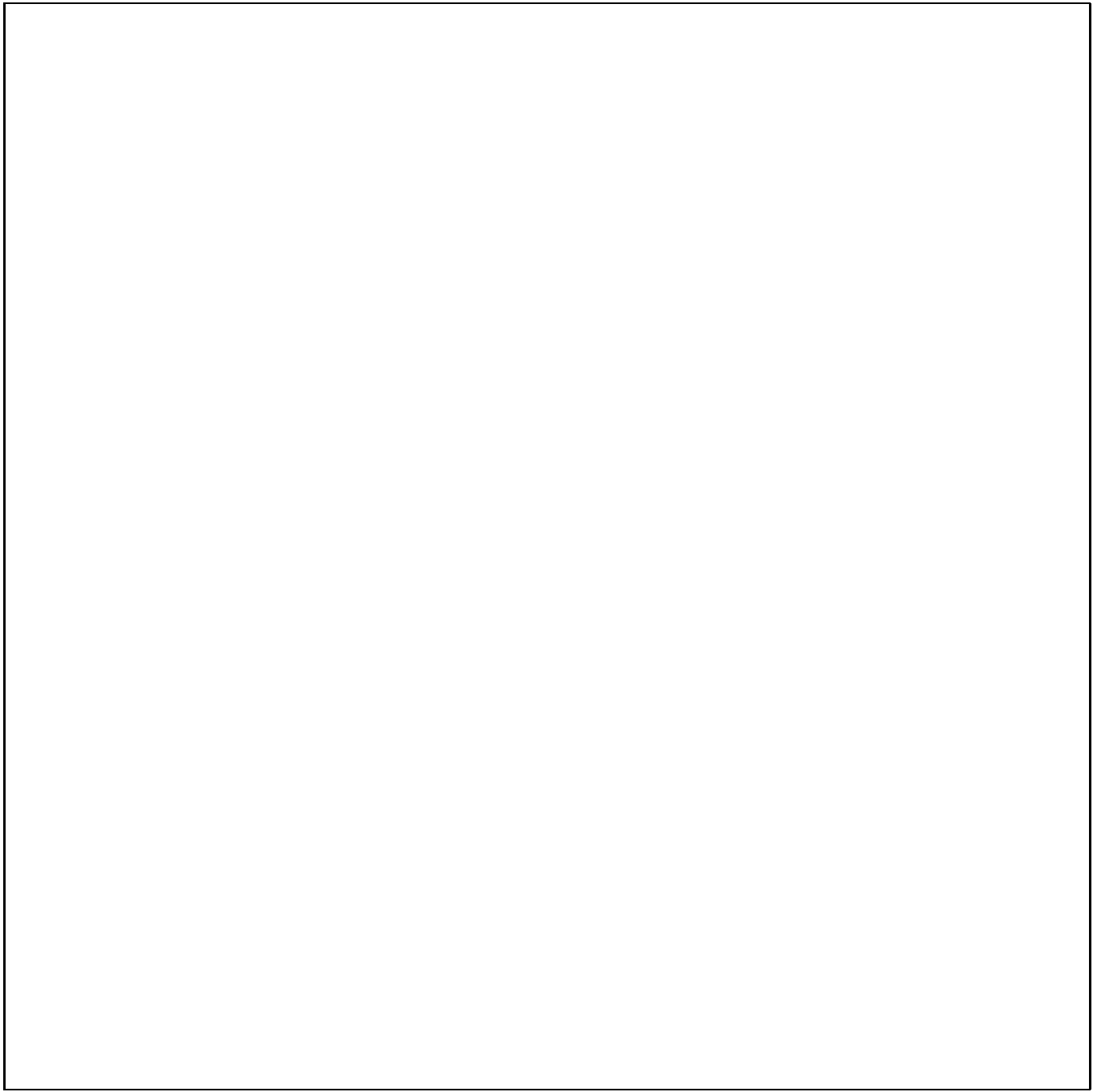

\picplace{18cm}
\caption[3]{ H$\alpha$  map of the IZW18 with 6600\AA\ red continuum contours 
overlaid.}
\end{figure*}  

         Figure 3 shows the H$\alpha$ image of the galaxy with the 6600~\AA\ 
red continuum contours overlaid. One sees that the  centroid of
the NW  H$\alpha$ emission is located in the plateau, between the NW
and (b) red continuum  peaks and  is displaced to the north of 
at 1\arcsec.5 (72pc) from NW red continuum peak. As it  was mentioned above 
(Sec. 3.1) 
the NW red component lies on an isolated stellar cluster-HII region complex
 (HT95). 
The (b) red peak is located on the NW H$\alpha$ emission region and its 
connection with this star--forming region is more evident. The centroid of 
the SE compact H$\alpha$ component is displaced by  1\arcsec.0 (48pc) to the 
west of 
the SE red continuum centroid. This displacement between the
continuum and H$\alpha$ images is  also evident from Fig.17 of HT95.

         The H$\alpha$ emission extends towards the NE and SW directions from
the main body of IZW18 up to distance of about 1.2 kpc. The same order of 
extension is detected by DH90 and M96. As we stress in 
Sec.3.1 in the same direction but for smaller distances the clumpy red 
continuum emission was detected. The extended envelope detected in 
H$\alpha$ line is not simply diffuse and shows numerous condensations (DH90). 

         We suggest that the isolated condensations that have been
 identified in 
the extended envelope of IZW18 as well as in the ridge are bonafide HII
 regions.
This  suggestion is supported by:\\
    - the presence of clumpy red continuum radiation of stellar origin  
      in the same fields where HII regions were identified.

    - the detection of  individual stars in the vicinity of the HII regions 
      fields (Table 2 of HT95).
 
    - the existence of nearly 15 well isolated HII regions in the main 
      body of the galaxy also reported by HT95. 
                 
         Since the  H$\alpha$ luminosity of any individual condensation is low 
(see below) the spectral type of the ionizing stars cannot be much earlier 
that B0 (Osterbrock 1989). At the distance of IZW18 such a star has a U 
magnitude equal to 25.4 which is roughly the limiting magnitude for the
 F336W frame of 
HT95. Nevertheless we carefully examined the positions of each individual HII 
regions to find such stars using the F336W frames of the   HST--WFPC2
 kindly given to 
us by Deidre Hunter. No such stars were found.
     
\begin{table*}
\caption[1]{HII regions in IZW18}
\begin{flushleft}
\begin{tabular}{cccccc}
\hline\noalign{\smallskip} 
\# & $\Delta$$\alpha $ & $\Delta$$\delta$ & $D$ & $F(H\alpha)$ & $L(H\alpha)$ \\     
 \small{} & \small{$\arcsec$} & \small{$\arcsec$} & \small{$pc$} & \small{$10^{-16} \ ergs \ cm^{-2}\rm s^{-1}$} & 
\small{$10^{36} \ ergs~\rm s^{-1}$} \\
\hline\noalign{\smallskip}
 1   &  10.9  &  -2.2   &  56  &   0.65   &  0.78 \\     
 2   &  10.2  &   0.0   &  97  &   1.75   &  2.11 \\    
 3   &  10.2  &   4.4   &  97  &   6.57   &  7.89 \\
 4   &   8.7  &   0.7   &  40  &   0.77   &  0.92 \\    
 5   &   6.5  &  15.2   &  56  &   1.68   &  2.01 \\   
 6   &   4.4  & -10.9   &  56  &   0.54   &  0.64 \\
 7   &   3.6  & -11.6   &  40  &   0.65   &  0.78 \\    
 8   &   3.6  &   5.1   &  56  &   6.27   &  7.52 \\ 
 9   &   3.6  &  13.1   &  97  &   2.67   &  3.21 \\ 
10   &   2.9  &  18.9   &  69  &   2.71   &  3.25 \\ 
11   &   2.2  &   6.5   &  40  &   3.48   &  4.17 \\
12   &   2.2  &  20.3   &  56  &   2.37   &  2.84 \\
13   &   1.5  & -16.7   &  97  &   4.55   &  5.46 \\  
14   &   1.5  &  18.9   &  40  &   1.68   &  2.01 \\  
15   &   0.7  & -13.8   &  40  &   1.49   &  1.79 \\
16   &   0.0  &  -8.7   &  40  &   6.69   &  8.03 \\
17   &   0.0  &  15.2   &  56  &   2.06   &  2.47 \\  
18   &  -0.7  & -14.5   &  79  &   3.97   &  4.77 \\ 
19   &  -0.7  & -10.9   &  69  &   7.03   &  8.44 \\ 
20   &  -1.5  &  10.9   & 112  &   7.19   &  8.62 \\  
21   &  -1.5  &  13.1   &  69  &   1.75   &  2.11 \\  
22   &  -2.2  & -17.4   &  56  &   2.79   &  3.35 \\   
23   &  -2.9  & -13.8   &  97  &   9.48   & 11.30 \\   
24   &  88.0  &  -7.3   &  56  &   6.80   &  8.17 \\ 
25   &  -5.8  & -15.2   &  40  &   0.92   &  1.10 \\   
26   &  -6.5  &  -4.4   &  56  &   7.15   &  8.58 \\  
27   &  -7.3  &   3.6   &  40  &   3.02   &  3.62 \\  
28   &  -8.0  &   4.4   &  56  &   3.86   &  4.63 \\ 
29   &  -9.4  &  -5.1   &  97  &  17.00   & 20.40 \\ 
30   &  -10.2 &  -4.4   & 112  &  19.20   & 23.00 \\   
31   &  -10.9 &   1.5   &  56  &   1.64   &  1.97 \\   
32   &  -12.3 &  -1.5   &  40  &   1.91   &  2.29 \\  
33   &  -13.1 &   0.0   &  79  &   7.53   &  9.04 \\    
34   &  -14.5 &   4.4   &  97  &   5.58   &  6.70 \\  
35   &  -14.5 &  -2.2   &  56  &   6.92   &  8.30 \\    
36   &  -16.0 &  -13.1  &  97  &   5.73   &  6.88 \\   
37   &  -18.1 &   -8.7  &  56  &   1.30   &  1.56 \\  
38(C)&  -21.1 &   17.0  &  63  &   0.77   &  0.92 \\   
39   &  -19.6 &  -13.1  &  40  &   0.80   &  0.96 \\ 
\noalign{\smallskip}
\hline
\end{tabular}
\end{flushleft}

\end{table*}
  
         A total of 39 HII regions were identified. They have been detected 
on the total H$\alpha$ map and their presence on at least two consecutive 
channel maps has been carefully checked. A majority of them are  well seen
on
 Fig.1 
and 2 of Ostlin et al. (1996) and also on  Fig. 1 of HT95. Observed HII
 regions 
are shown in the identification chart (Fig. 2) and listed in Table 2 which 
gives the following informations:

         Column 1 lists the reference number given to individual HII
regions as shown in the chart.

         Columns 2 and 3 give the X,Y positions of the photometric
center of the regions, in units of arcsec. The coordinate center
is  the  H$\alpha$ brightness peak  in the NW component of
the galaxy, positive values of $\Delta$$\alpha$ and $\Delta$$\delta$ are to 
the east and north respectively. The faintest H$\alpha$ fluxes correspond to a
S/N threshold of 5.

         Columns 4 and 5 give the effective diameter $D$ of the regions
in arcsec and parsecs. An equivalent diameter was defined in terms of the 
limiting isophotal area $A$, of each HII region, according to the definition
 
                          $D=2(A/\pi)\rm^{1/2}$

         The limits of each region are defined by the contour where
the intensity of the H$\alpha$ emission falls down to the average local
intensity of the diffuse background.

         Columns 6 and 7 list the total dereddened H$\alpha$ fluxes of the
regions in units of 10$^{-16}$ergs~$\rm cm^{-2}~s^{-1}$  and their H$\alpha$ 
luminosities in units of 10$^{36}$ergs~$\rm s^{-1}$.

	 The faintest features listed in Table 2 have an H$\alpha$ flux
corresponding to a minimum number of photons that gives a signal-to-noise of
5.
         On Fig. 2 is marked and listed in Table 2 the Zwicky's "flare" 
(Zwicky 1966), component C by DH90. About 1.5 arcsec displacement exists 
between its red (Table 1) and H$\alpha$ (Table 2) peaks. This displacement
 also appears on Fig. 2 of Ostlin et al. (1996).  
  
\subsection{HII regions size distribution and luminosity function}

\begin{figure}
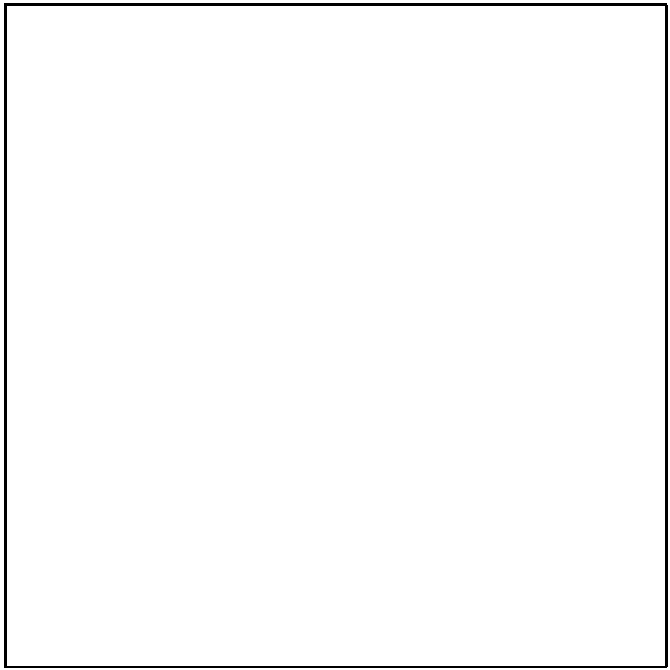

\picplace{8.8cm}
\caption[4]{Cumulative effective diameter distribution for the outer HII 
regions of IZW18.}
\end{figure}   

         In Fig. 4 we show the cumulative diameter distribution built
according to Table 2. It is consistent with an exponential law
$N$ =$N_{\rm 0}$exp($D$/$D_{\rm 0}$) 
where $D_{\rm 0}$ = 24 pc. This value is  comparable with
$D_{\rm 0}$ values 
found for the Irregular galaxies (Hodge 1983; Hodge et al. 1989; 
Hodge \& Lee 1990).

\begin{figure}
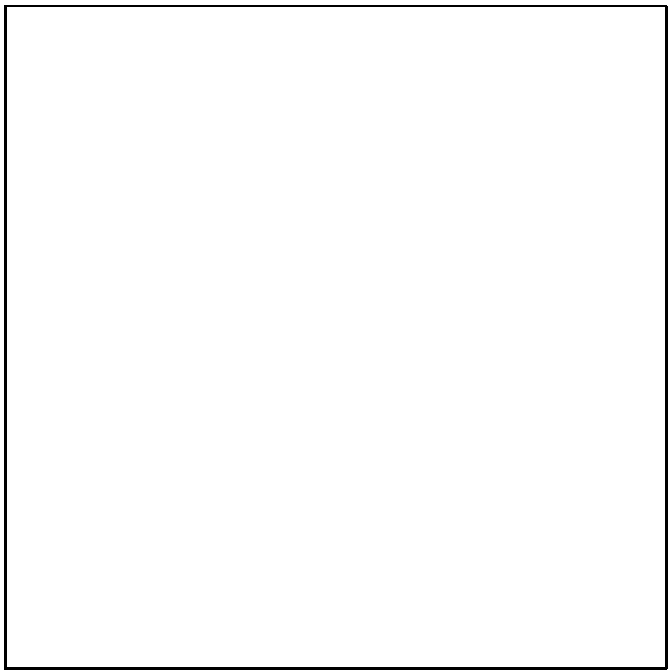

\picplace{8.8cm}
\caption[5]{The H$\alpha$ luminosity function for outer HII regions of IZW18.}
\end{figure}

         For IZW18 the small number of HII regions prevents a
derivation of an accurate luminosity function, but nevertheless
the data are sufficient to get a first approximation. In Fig. 5
the distribution of H$\alpha$  luminosities binned in intervals of 1.0
 10$^{36}$ergs~$\rm s^{-1}$ is presented. The observed luminosity function of
IZW18 shows a turnover with a maximum at $L$ = 3.0 10$^{36}$ergs~$\rm
s^{-1}$  corresponding  to the luminosity of the Orion nebula (Gebel 1968).
 This turnover is due to incompleteness  towards fainter luminosities. 
At luminosities brighter than this turnoff the observed function is 
reasonably well fitted by a power law $N(L)$=$A$$L^\alpha$$dL$ with a slope   
$\alpha$ = -1.6$\pm$0.3. This is consistent with the mean value derived for dwarf 
irregular galaxies (Strobel et al. 1991) $\alpha$ = - 1.5$\pm$0.3. It may be
compared with the steeper slope  $\alpha$ = -1.7 found for larger
irregulars by Kennicutt et al. (1989).

\subsection{H$\alpha$ velocity field} 
         Figure 6  shows iso-velocity lines superimposed on the H$\alpha$ 
gray scale image of IZW18. Figure 7 is a position-velocity cut across the 
optical major axis of the galaxy across the NW and SE compact components 
(PA = 143$^{\degr}$). The shape of this  position-velocity cut is in good
agreement with 
slit 2 (PA = 156.1$^{\degr}$) position-velocity diagram of M96.

\begin{figure}
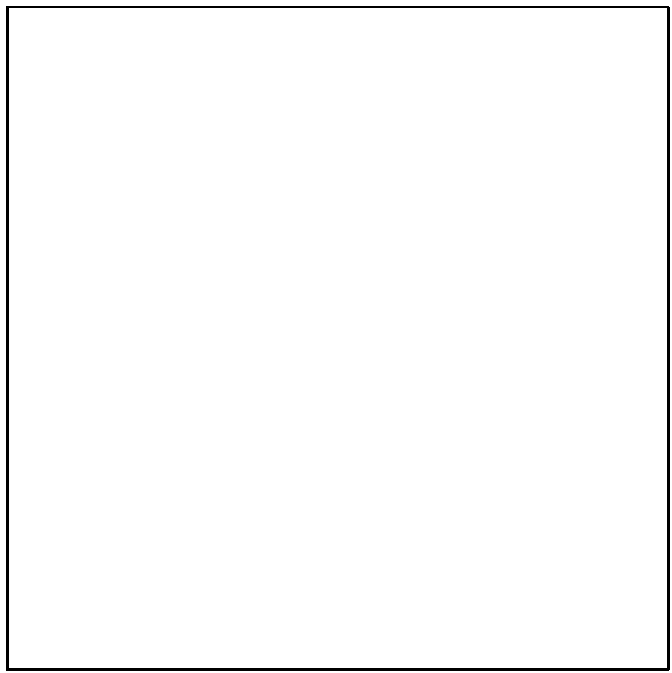

\picplace{8.8cm}
\caption[6]{ H$\alpha$ iso-velocity contour map of IZW18
 superimposed on a gray scale image of the galaxy.}
\end{figure}

\begin{figure}
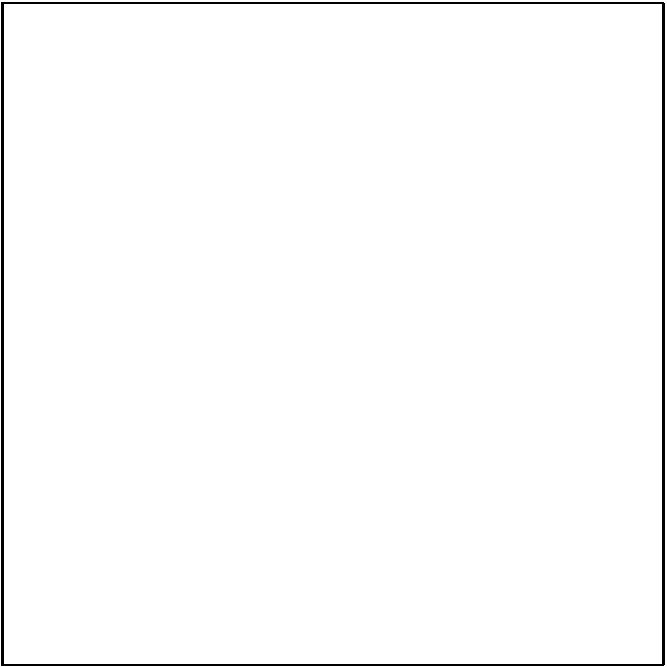

\picplace{8.8cm}
\caption[7]{ H$\alpha$ position-velocity cut across a line joining
 the NW and SE main compact HII components.}
\end{figure}

         Figures 6 and 7 show a clear overall regular
 radial velocity gradient across the main body of IZW18. This finding obtained
 from the ionized gas confirms an earlier result by Viallefond 
et al. 1987 using HI aperture synthesis techniques. Fig. 7 further shows that
the radial velocity gradient is globally consistent with a solid-body-like
rotation of the ionized gas.
         The peaks of H$\alpha$ emission in the NW and SE components are
respectively centered at 739~km~$\rm s^{-1}$  and 779~km~$\rm s^{-1}$  and 
their average velocities are equal to 742$\pm$7~km~$\rm s^{-1}$ and 
783$\pm$5~km~$\rm s^{-1}$. These values are in agreement with 
those obtained using high-resolution spectroscopy (DKF89). They 
are also very similar to the Davidson \& Kinman (1985) and HI velocity 
results (Lequeux \& Viallefond 1980; Viallefond et al. 1987).
         The best symmetrical pattern of the velocity distribution is obtained
 along  the  major axis, when  
assuming  a systemic  recession  velocity  of 763~km~$\rm s^{-1}$. 
This value is in good agreement with the velocity of the number 5 HI cloud 
(765~km~$\rm s^{-1}$) in Lequeux \& Viallefond 
(1980) and  relates to a region which is very close to the 
red continuum component (a).
         Assuming that the radial velocity gradient, (which amounts to 
73~km~$\rm s^{-1} \  kpc^{-1}$ and is  indicating a high mass concentration),
 is due to rotation (M96) we derive a Keplerian spherical mass of 
$1.9 10^{8}M_{\odot}$ within a radius about 10 arcsec (480pc). Within 
this radius, we derive $M(HII) = 3.1 \ 10^{6}M_{\odot}$ for the mass of 
ionized gas. It corresponds to  1.6\% of the total mass and about 
40\% of the neutral HI mass of the clump corresponding to the main body 
of IZW18 (Lequeux \& Viallefond 1980; Viallefond et al. 1987). Within this
limited radius, such a simple approximation is sufficient to derive a realistic
estimate of the total mass. Any other more elaborate model will provide an
estinate within a factor of 2 as compared to the Keplerian spherical mass. 

 Over the regular velocity field  of IZW18 
some complex 
structure is superimposed that locally modifies the isovelocity line and
creates "bumps" at the position-velocity cut. These irregularities are
particularly conspicuous in the NW region. Across a 70x70 pc$\rm ^{2}$  
 area located 150pc north from the peak intensity of the
NW component  we observe a local velocity excess of  
15-17~km~$\rm s^{-1}$. The position of this  excess coincides 
with the secondary peak in the spectrum of DFK89. It is also coincident
 with 
the location of the red continuum (b) component (Fig. 1, insert).
This velocity structure has been previously discussed by M96 and by 
Skillman \& Kennicutt (1993). The latter authors suggest a merging of two 
(or more) clouds rather than a pure solid body rotation. We dispute this
interpretation using our previous statements on the regularity of the general
velocity pattern and further note that the violent star formation that
occurs in the NW compact component is likely to induce local peculiar gas
motions.
         Other interesting velocity features relate to the ridge. In
 general
 the 
ridge follows  the velocity distribution of the HI peak and IZW18 
velocity field. Along the southern and part of the western length of the ridge,
differences in radial velocities of the order of 10 - 15 ~km~$\rm s^{-1}$ 
are found between its inner and outer regions. This difference is also evident 
across the slit 3 (PA=78.8$^{\degr}$) position-velocity diagram of M96. 
No peculiar velocities were detected on the  HII regions area of the 
ridge.

         Coincident with Zwicky's "flare" a small H$\alpha$ knot has a radial
velocity of about 730 ~km~$\rm s^{-1}$. Therefore IZW18 and the
 Zwicky's "flare"  
belong to the same system. Those HII regions detected in the main body of
IZW18 
and its envelope (Sec.3.2) with sufficient S/N ratio  exhibit 
radial velocities similar to that of the central velocity field.

\subsection{Velocity dispersion distribution}

\begin{figure*}
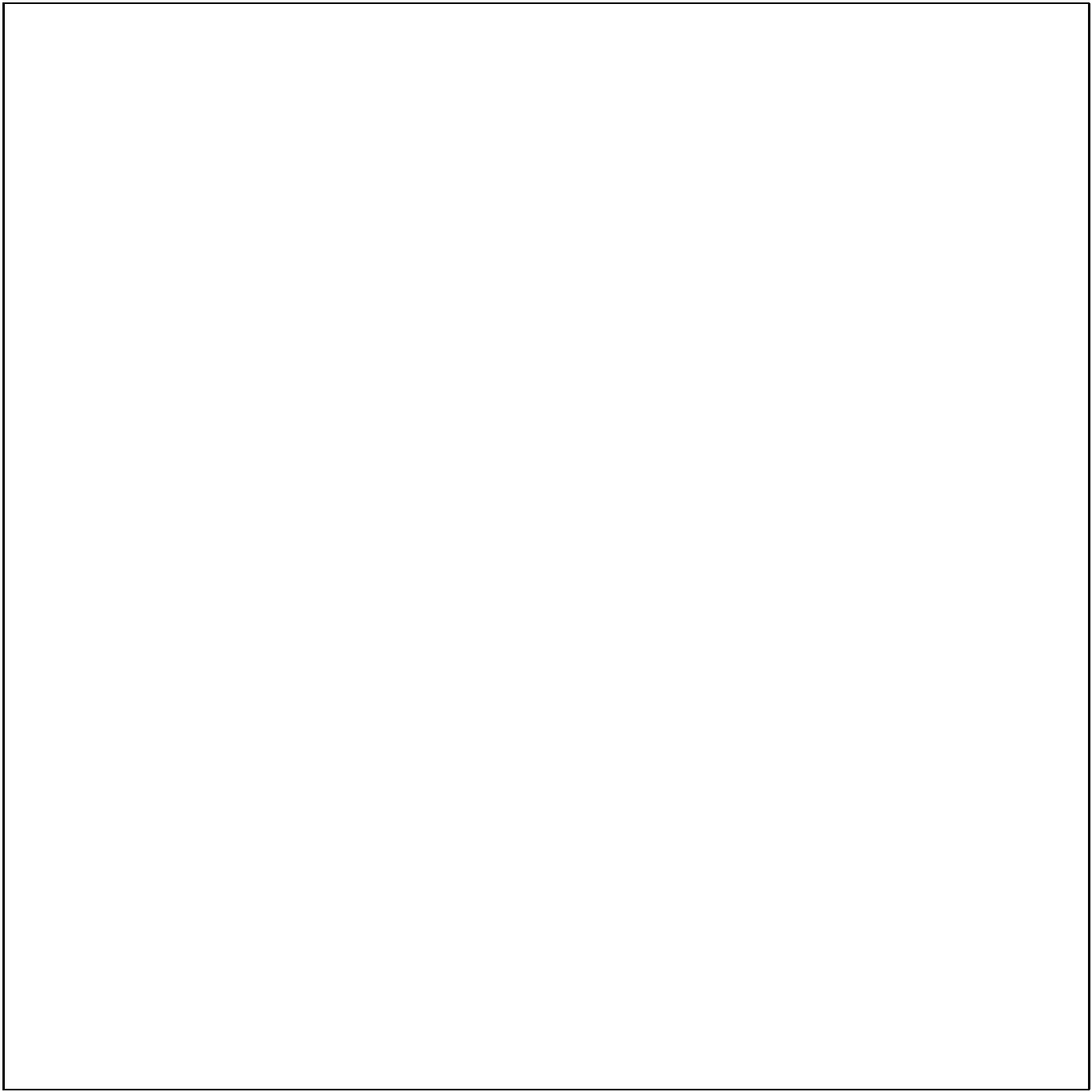

\picplace{18cm}
\caption[8]{H$\alpha$ line profiles on IZW18. Each profile is normalized 
to the maximum intensity of each pixel. The horizontal axis of 
each box represents the free spectral range of 375 ~km~$\rm s^{-1}$.
 In the chart, south 
is in the top and east is in the right side.} 
\end{figure*}

\begin{figure*}
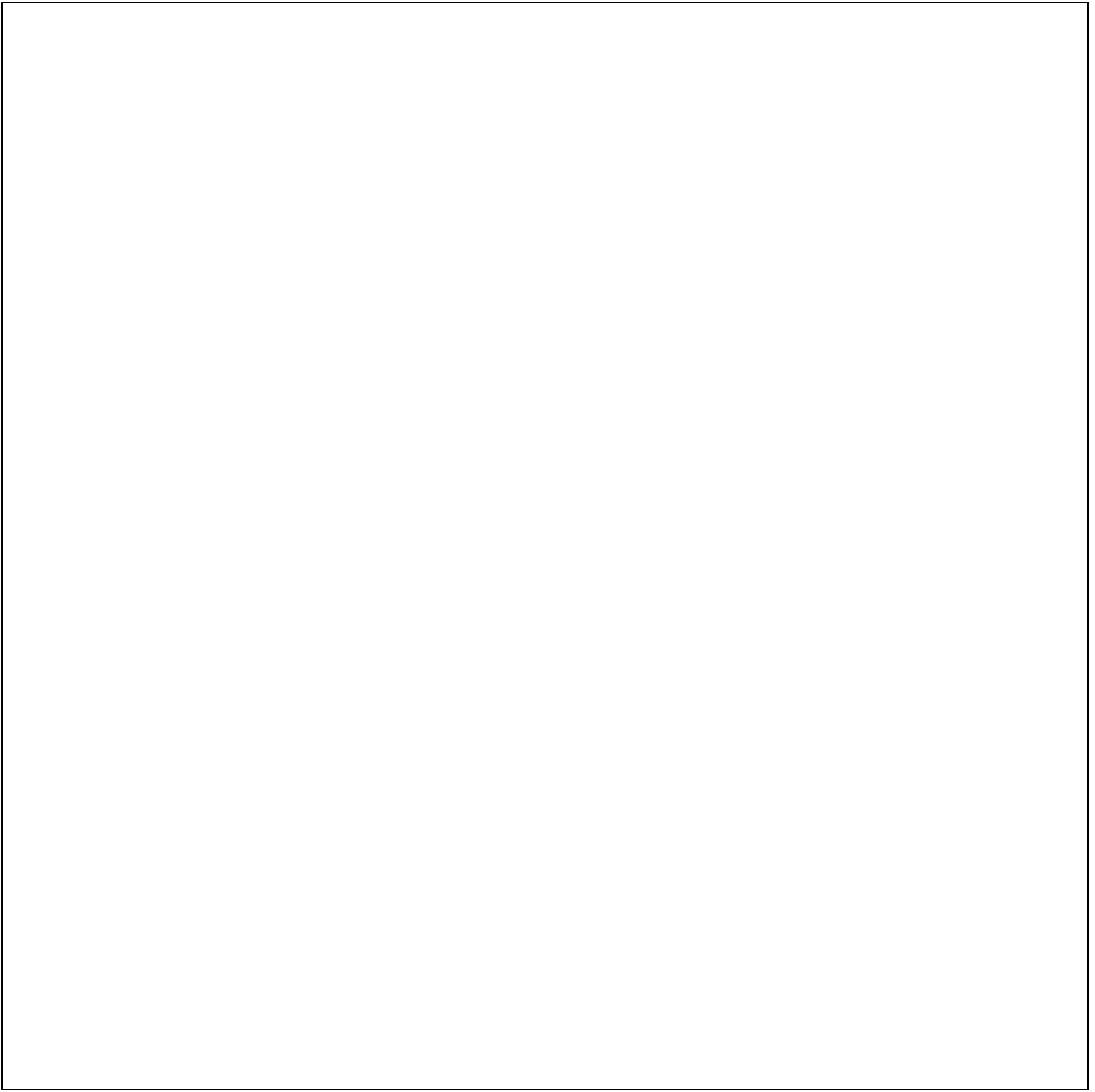

\picplace{18cm}
\caption[9]{The same as in Fig. 8 but each profile is normalized 
to the maximum intensity of the brightest pixel.}
\end{figure*}

         Figure 8 and 9  display each individual H$\alpha$ line profile
 superimposed to the H$\alpha$ isophotal map of IZW18  in order to
 show the 
line profile structure of the outer envelope and the H$\alpha$
line intensity distribution across the galaxy. Depending upon their location
 line profiles show different shapes. Typical unsmoothed
H$\alpha$ line profiles  are shown in Fig. 10 together with the corresponding
Gaussian fits. In all cases, the observed  broad wings 
have instrumental origin and  are most probably caused by adjacent 
orders.

         Table 3 reports the results of the line profile analysis
conducted on areas where the signal-to noise ratio is sufficient
to achieve gaussian or multi-gaussian fitting of the observed H$\alpha$ 
line.   $\beta$(main) is the velocity dispersion of the main component of 
the line , corrected for thermal and instrumental broadening (Sec. 2); 
n(sec) is the number of
secondary gaussian profiles fitting the observed H$\alpha$ line; 
V(sec) is the peak velocity difference between the  main and secondary
gaussian components; I(sec)/I(main) is the intensity ratio of the
secondary and main gaussian components. The numbers in Table 3
are average values taking into account several pixels for each
region.

\begin{table*}
\caption[1]{H$\alpha$ line profile analysis}
\begin{flushleft}
\begin{tabular}{cccccc}
\hline\noalign{\smallskip}                              
  &    &        NW region & SE region &  W ridge  & S ridge \\  
\hline\noalign{\smallskip}                                      
$\beta \ (main)\ (km \ s^{-1})$  &   & $35.3 \pm 1.6$ & 26.5$\pm$0.4  & 16 & 16 \\ 
n(sec)                      &   &     0        &     1         &  2 &  2 \\
V(sec)(km s$^{-1}$)   & $1:$  &   &-53.0$\pm$2.7 & 45.3$\pm$11.9 &-40.7$\pm$3.2 \\ 
     &                $2:$  &   &           & -39.0$\pm$7.0 & -76.8$\pm$5.1 \\
 I(sec)/I(main)     & $1:$  &   &    0.15      &    0.30       &    0.40 \\
     &                $2:$  &   &             &    0.20       &    0.20   \\ 
\noalign{\smallskip}
\hline
\end{tabular}
\end{flushleft}
\end{table*}

         The NW component H$\alpha$ line profiles mainly appear 
symmetrical and well-represented by a one-component Gaussian fit 
(Fig. 10a). It is interesting to note that in H$\alpha$ line a small red 
asymmetry appears towards the direction of the SE component. With the 
assumption that the component is a gravitationally bound system 
(Terlevich \& Melnick 1981) the total mass of the system is estimated 
about 2 10$^{7}$$M_{\odot}$ and the fraction of ionized gas in it as about 3\%.

          According to Skillman \& Kennicutt (1993) a broad emission can be
seen at the base of the H$\alpha$ line profile for the NW component with a 
peak amplitude at about 0.2\% of the main H$\alpha$ line and
a FWHM of about 80\AA\ (about 3600~km~$\rm s^{-1}$). The small free spectral
range (375~km~$\rm s^{-1}$) of the Fabry-Perot interferometer prevents
to detect this broad component. 

\begin{figure*}
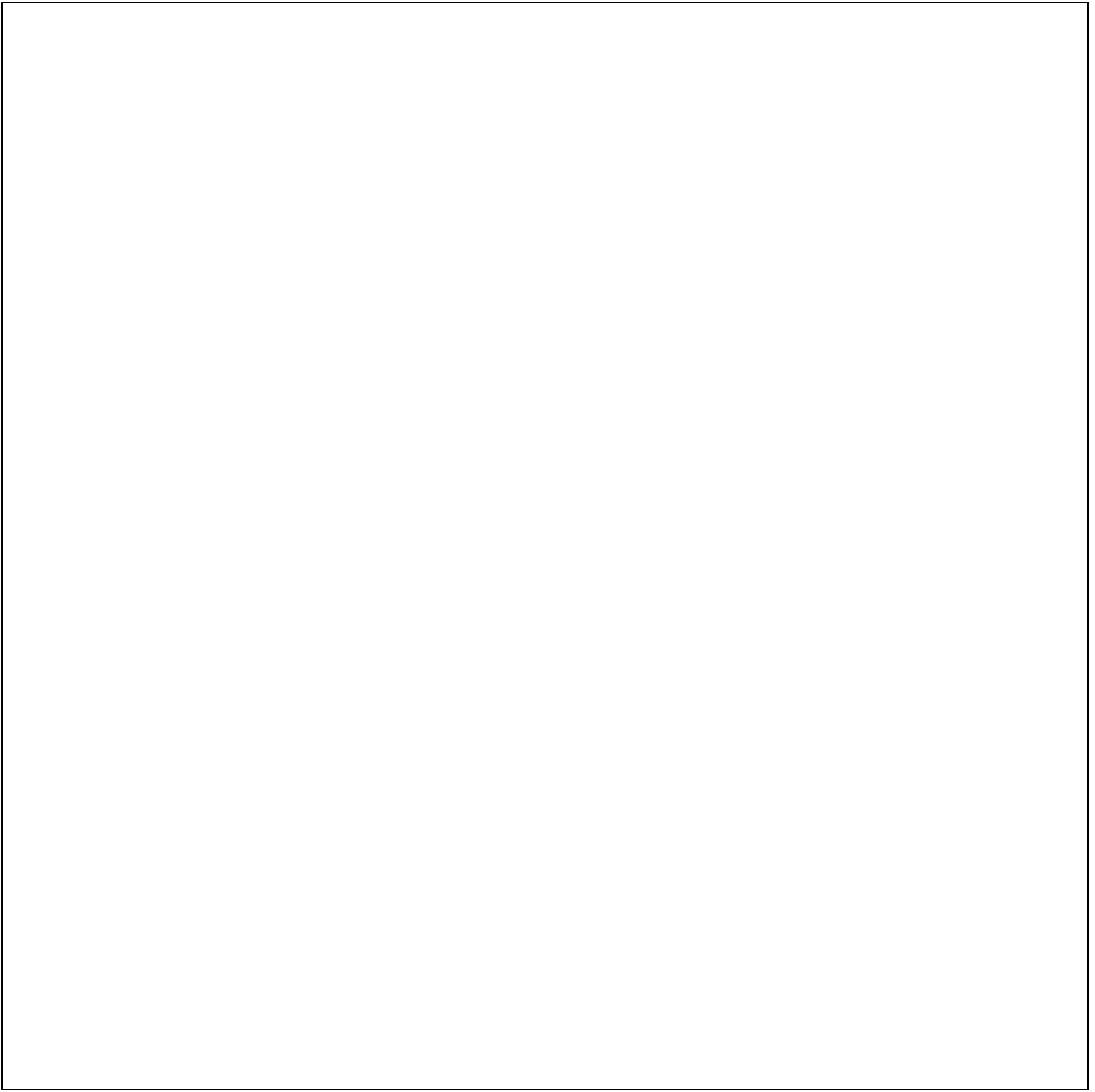

\picplace{18cm}
\caption[10]{Observed H$\alpha$ profiles typical for: (a) NW and (b) SE 
components 
of IZW18; (c) for western part of the ridge and (d) southern part of the 
ridge. Broken lines are the fitted gaussian profiles. Ticks in velocity
scale are separated by 17~km~$\rm s^{-1}$.  }
\end{figure*}       

          All H$\alpha$ line profiles observed in pixels belonging to 
the SE component are asymmetric with blue side excesses (Fig. 10b). This
blueshifted component contributes in average to 13\% of the line flux and has
a mean velocity of -50~km~$\rm s^{-1}$ from the systemic velocity of the main
component. Again, the small free spectral range of our observations
 (375~km~$\rm s^{-1}$) forbids the detection of faint large velocity
components such as the ones found by M96 with velocities up to
 230~km~$\rm s^{-1}$ and A an intensity of 
about 3\% of the main line flux.   

         Across the western part of the ridge (Fig. 10c)  
profiles with blue and red velocities excesses are typical. Usually 
redshifted components are more intense than blueshifted ones. 

         Across the southern part of the ridge profiles with strong blue 
side asymmetry are typical (Fig. 10d). Two secondary components with 25\% and 
12\% 
contributions of the line flux and -41~km~$\rm s^{-1}$  and 
-77~km~$\rm s^{-1}$  blueshifted from the systemic velocity of the main 
component were discovered. The bright knot  well seen on the M96 Echelle
spectrum at H$\alpha$  along the slit
position 1 (PA=7.7$^{\degr}$)  (see her Fig. 2)  is caused 
by the southern part of the ridge. The two blueshifted components in the
H$\alpha$ profiles of the ridge are well separated in velocity (see
Fig. 10d). This is well in  agreement with the velocity range displayed
in the M96  spectrum observed across the 
ridge position. On the so--called SW Doppler ellipse (M96), nearly 4\arcsec
south from 
the ridge position, another fainter knot is observed that refers to our
 HII region number 23.      
    
         In the outer enveloppe of the main body of the galaxy, profiles 
 are mainly asymmetric with red wings in the north-west and  west of the SE
component (M96), and with blue wings elsewere. In the NE Doppler ellipse of 
M96 (her Fig. 2), three other  HII regions from our list have been
identified as they appear as individual knots: number 9 and number 10 merged with number 12. All three HII
 regions have
H$\alpha$ line profiles with blueshifted components.        

\section{Discussion}

\subsection{The structure}

         The question still remains whether IZW18 is an old system in which
 some intense episodes of
star formation already occured, or whether it forms stars for the very
first time? Loose \& Thuan (1986) reported observations of a faint
outer component with an extended stellar continuum which they
attributed to an underlying old population. Thuan (1983) detected
infrared emission, while Hua et al. (1987), DFK89, DH90 from red continuum 
imagery, confirmed the presence of this component. The implication is 
that IZW18 may be an older and more complex system.

         In the following, we make the assumption that H$\alpha$ peaks 
relate to the current star--forming regions while red continuum
features correspond
to older sites of star formation. From the existence of the H$\alpha$ loops
 and filaments HT95 also
suggested  the existence of intermitent star formation episodes 
in the central and northern parts of the galaxy. Their  positions are
in good agreement with our (a) and (b) red components. 

         The distribution of the clumps in the main body of IZW18 shows a 
chain--like structure. Out of seven identified current or more evolved
 star-forming knots 
six are distributed along the line SE,(a),NW,(b). This chain of star--forming
 knots is positioned 
along the 
direction of elongation of the main body of IZW18 and is
 consistent with HI outer clumps distribution  
(Lequeux \& Viallefond 1980). Only knot (c) is displaced from this linear 
distribution. 

         In the chain both NW and SE currently active star--forming regions are
located between more evolved star--forming sites. Is this
geometry a random configuration?  Does it examplify the sequential
nature of star formation events in the galaxy (Gerola et al.
1980; Kunth et al. 1988) ? Clearly the chain although well ordered
spatially does not define any prefered age sequence .
 Stochastic self-propagating star
formation simulations of star formation sites in dwarf galaxies
show that the properties of dwarf systems  depend on the size ratio of the
 galaxy to the star
formation cells. Small galaxies similar to IZW18 evolve
mainly via a series of disconnected star forming bursts (e.g.
Hunter \& Gallagher 1985). Therefore, a stochastic  origin of the observed
 "chain" 
is more plausible (see also HT95).

         The ridge discovered by DH90 is thought 
to represent a radiation-bounded ionization front  driven into the 
main HI cloud.  Our red continuum data  dispute this claim. 
Indeed the
discovery of some stars in the ridge and surrounding regions (HT95) shows
 that the ridge contains a stellar population that can ionize the gas ``in
situ''. Therefore the ridge is  an isolated structure in the W-SW part of 
the double-peaked HI gas condensation and  has no direct morphological link 
with the  main body of IZW18.  
 
         Extended H$\alpha$ emission have been detected up to distances of
 1.5  kpc from the main body of IZW18. DH90, who noted the 
clumpy nature of this extended H$\alpha$ emission,
 suggested that this emission could 
be powered by the central NW and SE  star--forming complexes. M96 using
 imagery and
 high dispersion spectroscopy advocate that both north-east and 
south-west extensions of the outer envelope of the galaxy could  be 
superbubbles powered by  star--forming complexes. However the existence 
of  clumpy  red continuum  emission of stellar origin coeval
 with the extended 
H$\alpha$ emission,
as well as the  number of stars which were discovered 
in the same regions (HT95) consisting into  compact and well 
individualized clumpy structure  allow us to suggest that 
a stellar population is embedded into the ionized gas. 
Therefore part of the young stars from this population can ionize
 and produce the observed extended HII emission.  As noted in Sec.3.3, the
 diameters distribution and a 
tentative H$\alpha$ luminosity function of these HII regions are consistent 
with those of normal HII regions in dwarf irregular galaxies. The existence 
of the stellar and HII regions  populations in NE and SW extensions of the 
outer envelope of IZW18 does not preclude the contribution  of some 
superbubbles in the same regions.       

\subsection{The kinematics and dynamics}

         As previously suggested, star formation in the main body
of IZW18 possibly occurs via a series of disconnected bursts as
a result of density fluctuations in the interstellar medium near
the HI clouds. What is the mechanism which triggers the subsequent 
star formation? The velocity field in an HII galaxy as well as in its 
surrounding HI cloud may indeed play an important role (Saito et al. 1992; 
Tomita et al. 1993).

         The velocity gradient observed along the main star--forming regions,
both in H$\alpha$ (our results in Sec.3.4) and in HI (from the maps of 
Viallefond et al. 1987) is equal to about  70 ~km~$\rm s^{-1} kpc^{-1}$. The
same gradient across the offseted HI peak distribution is only 
50 ~km~$\rm s^{-1} kpc^{-1}$ (Viallefond et al. 1987). It means that local HII
 as well as HI gas in the region of the optical galaxy rotates faster
 than the HI gas at
 its peak distribution.  

          The ordered component of the IZW18 velocity field  shows a peak
orbital velocity of about 50 ~km~$\rm s^{-1}$. Most irregular galaxies 
(giants as well as dwarfs), exhibit a solid body rotation with about the
same peak 
orbital velocities of 50--70 ~km~$\rm s^{-1}$ but with shallow velocity 
gradients of 5--20 ~km~$\rm s^{-1} kpc^{-1}$ over the optically visible 
region (Hopp \& Schulte-Ladbeck 1991; Shostak \& Skillman 1989; Gallagher 
\& Hunter 1984 and literature therein). The much higher velocity gradient 
in IZW18 arises from its much smaller linear scale, when compared to 
normal irregulars.

          The disordered motions superimposed over the regular
velocity field of IZW18 appear as sharp velocity jumps of  
10 to 20 ~km~$\rm s^{-1}$ occuring on scales of about 500 pc$\rm^{2}$. These
appear related to the main body of the galaxy, and suggest that
the gas is not in equilibrum locally. Corresponding dynamical
timescales could be of the order of few 10$^{6}$years (HT95). The
gravitational effects of the (b) and (c)  star--forming centers as well as
explosive star--forming events  may explain these local
disturbances.

         As regards to H$\alpha$ line profiles, the discrepancy
between a single  Gaussian component with velocity dispersion of
23 ~km~$\rm s^{-1}$ (e - folding width: 33 ~km~$\rm s^{-1}$) observed 
in a 40 arcsec and 21 arcsec apertures by Arsenault \& Roy (1986) and the 
complex line structure seen at our resolution of 0.\arcsec725 per pixel
is not a surprise. Gaussian-like smooth profiles obtained by
Arsenault \& Roy (1986)  result from integration
over many distinct components with different space velocities and
emissivities. As examples we can recall high resolution
observations of NGC604 by Rosa \& Solf (1984), Hippelein \& Fried
(1984) in comparison with results for NGC604 presented by Arsenault \& Roy
 (1986) and 
also the very interesting case of the giant HII region NGC2363 (Arsenault
\& Roy 1986; Roy et al. 1991; Roy et al. 1992).

         The single  Gaussian component  observed in the
NW compact HII region may suggest  a self-
gravitating system (Terlevich \& Melnick 1981). In this case the
expected velocity dispersion (Melnick et al. 1988) is about twice
smaller than the observed one. This difference can be caused either
by the gravity of the nearby located recent star--forming NW and
(b) regions or/and the effects of turbulence. The chaotic morphology of the
ionized gas in the component (HT95) as well as the existence 
of high velocity gas expanding shell (M96) may favor the latter effects.
 Application 
of gravitation - turbulence model in this case would be more reliable 
(Arsenault \& Roy 1988).

          In the SE compact HII region the velocity dispersion
for the main Gaussian component (about 19 ~km~$\rm s^{-1}$) is also higher than
 expected from a self-gravitational model (Terlevich \& Melnick
1981). Such an  excess velocity dispersion can also be caused by the 
peculiar motions of several stellar clusters observed in the core of this
component (HT95). The weaker second Gaussian component 
may be produced by another isolated HII region in the SE area,
 undistinguishable
because  of projection effects.
       
        The ridge which has no direct morphological connection with the main
body of IZW18 broadly follows the velocity field structure of the 
HI peak. The ridge is therefore co-rotating with the HI 
cloud and the main body of IZW18. The ridge as an isolated configuration 
is probably not in 
equilibrium. It expands in an  opposite direction with respect to the HI
gas  showing local  well determined infall and/or 
outfall of ionized gas. The geometry as well as the velocity field of the
 ridge are
consistent with the  "blister" interpretation .
  The complex structure of the H$\alpha$ line profiles observed
in the envelope showing a variable position-dependent asymmetry
 (blue and/or red
wings) suggests that we are observing gas infall onto the main
body of the galaxy, whatever the direction of observation and/or local
ionization fronts. The
blue asymmetric components are much more prominent on the eastern
side, opposite to the major HI concentrations. It may be evidence
that infall of gas clouds fuels the overall star formation in
IZW18. But we do not exclude the
 possibility that the 
 arc-like structure of the ridge can be affected by a tidal interaction 
between  IZW18 and the Zwicky's "flare".

\subsection{The nature of IZW18}

         From our data and data from the literature we examine
the nature of IZW18 following the hypothesis first introduced
by Searle \& Sargent (1972) and further discussed by  Kunth et al. (1986),
 Vigroux et
al. (1986), Thuan (1987) that in general
BCDGs may be considered as the low luminosity end of irregular
galaxies. In Table 4 global characteristics and parameters
related to star formation properties in IZW18 are compared with
those of dwarf irregular galaxies.

\begin{table*}
\caption[1]{IZW18 and Dwarf Irr properties}
\begin{flushleft}
\begin{tabular}{ccccc}
\hline\noalign{\smallskip}                              
    &   IZW18 & Ref  &   Dwarf Irr  &  Ref \\ 
\hline\noalign{\smallskip}                                  
M(B)             &        -13.9 & Huchra (1977)  &  -14.5    & Hunter \& Gallagher (1985, 1986)\\
$D \ (kpc)$           &          0.9 & Huchra (1977)  &    5.9    & Hunter \& Gallagher (1985, 1986)\\
$BSB(mag~arcsec^{-2}$) &      21.3 & Huchra (1977)  &   23.8    & Hunter \& Gallagher (1985, 1986) \\
$lgM(HI)_{\odot}$   &         7.8 & Lequeux \& Viallefond (1980) & 8.3    & Hunter \& Gallagher (1985, 1986) \\
$V(max)(km s^{-1}$) &          50 & This paper     &     60    & Gallagher \& Hunter (1984) \\ 
$dV/dr(km~s^{-1}~kpc^{-1}$) &   73 & This Paper     &     10    & Gallagher \& Hunter (1984) \\
U - B             &       -0.76 & Huchra (1977)  &  -0.44    & Hunter \& Gallagher (1985, 1986) \\
B - V             &        0.17 & Huchra (1977)  &   0.36    & Hunter \& Gallagher (1985, 1986)  \\
Largest HII reg.(pc) &      170 & This Paper     &    300    & Hunter \& Gallagher (1985) \\
$D_{0} \ (pc)$            &       24 & This Paper     &     25    & Hodge (1983) \\
$\alpha$               &     -1.6 & This Paper     &   -1.5    & Strobel et al. (1991) \\
$O/H(10^{4})$          &     0.17 & Dufour et al. (1988) &   4.43  & Hunter \& Gallagher (1985, 1986) \\
$L(H\alpha)(L_{\odot})$ & $3 10^{5}$ & This Paper    &  $3 10^{4}$ & Hunter \& Gallagher (1985) \\
$SB(H\alpha)(L_{\odot} pc^{-2})$ & 9 & This Paper &    1.5    &  Hunter \& Gallagher (1985) \\
$SFR(M_{\odot} yr^{-1} pc^{-2})$   & $4 \ 10^{-8}$ & This Paper  & $10^{-10}$ & Hunter \& Gallagher (1985)  \\ 
\noalign{\smallskip}
\hline
\end{tabular}
\end{flushleft}
\end{table*}  

The results of comparison may be summarized as follows.\\
1. IZW18 has an absolute luminosity consistent with that of an
"average" dwarf irregular. The small linear size of IZW18 accounts for
the high surface brightness of the optically visible component.

2. HI masses of IZW18 and dwarf irregular galaxies of comparable
blue luminosity and  the clumpy structure of the HI distribution
are similar. As for IZW18, star formation in dwarf irregulars is
not simply related to the projected HI gas density (e.g. Allsopp
1978).

3. A rigid body rotation which is typical for the majority of dwarf
irregular galaxies with similar peak orbital velocity is
observed in IZW18. The size of IZW18 makes the velocity gradient
across it at least 5 times higher than in irregular galaxies.

4. Except for two giant star--forming sites, IZW18 contains a population of
faint HII regions. This population by its luminosity and diameter distribution
does not differ from the HII region population in dwarf irregular galaxies.

5. In IZW18 besides an extended underlying older stellar
population (Loose \& Thuan 1986) we report the existence 
of recent  star--forming clumps. These regions also
are important components in dwarf irregular 
galaxies (Hunter \& Gallagher 1986).

6. By their very blue color, high H$\alpha$ luminosity and surface
brightness and their high star formation rate per unit area, the
current star--forming regions in IZW18  significantly differ from
those in dwarf irregular galaxies.

         We may conclude that IZW18 share many characteristics in
common with dwarf irregulars except for its much higher level of
star--forming activity per unit area. The large velocity gradient together
 with
 the locally complex velocity field (including peculiar gas motions
responsible for the line profile asymetries) are most probably
 responsible for the star
formation itself.

         From the existence of recent star--forming
regions seen in continuum, with no exact spatial coincidence with
HII regions, we can possibly conclude that we are observing a major
 second burst
of star formation in this object (e.g. Kunth et al. 1995). The first 
burst was completed about few tenth 10$\rm^{6}$ years ago (e.g. HT95; M96) and 
the present one is extremely recent (Kunth \& Sargent 1986). The age of the 
present burst may be estimated as no more than 
a few 10$\rm^{6}$ years (Lequeux et al. 1981; 
Copetti et al. 1985).

\section{Conclusions}          
         
         The main results of H$\alpha$ scanning Fabry-Perot investigations 
of IZW18 may be summarized as follows:\\
 - In the main body of this galaxy, the {\it current} sites of active
star formation, identified from the presence of prominent
 emission lines  have  ages of few 
10$^{6}$ years and are  spatially separated from red continuum clumps which 
are  sites of  recent star formation episodes with ages 
of about few tenth of 10$^{6}$ years. All star formation sites in the galaxy 
are separated from the compact HI cloud double-peaked core.

 - The extended H$\alpha$ emission seen in the envelope surrounding the
central condensations exhibits a number of clumps which are thought to be 
a population of normal HII regions.

 - The velocity field of the galaxy shows a regular pattern suggesting 
ordered solid-body rotation on which local disturbances are superimposed.

 - The H$\alpha$ line profiles have a complex structure, and except in
the NW HII compact region, cannot be fitted by single gaussian
components. The presence of blue and/or red wings in many points
of the envelope suggests a generalized pattern of infalling (expelling) gas
clouds onto (from) the central star--forming regions.

 - The observed "ridge" is an isolated morphological structure in the
system. It is a star--forming site separated with respect to the IZW18 
main  body by  the HI cloud. The shape and orientation of this
 ridge can be due to tidal interation between IZW18 and the Zwicky's "flare".

 - The SW and NE extensions of the outer envelope of the galaxy besides the
ionized gas component connected to a supergiant shell-expanding 
superbubble (M96) also  contain stellar as well as HII regions components. 
 
 - The near environment of the galaxy is likely to be  polluted not only with 
gas processed by the central massive stars and driven out by superwinds but
 also by products of a more local stellar population.
    
         When compared to the average characteristics of dwarf
irregular galaxies, the basic properties of IZW18 (luminosity,
HI mass, solid body rotation, normal HII region population,
existence of current and recent star formation sites) are
consistent with the hypothesis of an overall similarity between
this blue compact dwarf galaxy and a typical dwarf irregular. 
The rigid-body rotational velocity gradient and the blue surface
brightness, which depend on the linear scale are much
larger in IZW18. The essential difference is that a high star
formation activity (an order of two magnitudes in the SF rate per
unit area which respect to that observed in a typical dwarf
irregular) is concentrated in a small compact region.

         $Acknowledgments.$ {We would like to thank Dr. Deidre Hunter for 
providing her HST-WFPC2 data for IZW18 and Dr. Miguel Mas-Hesse for his model 
calculations of stellar contribution at 6600~\AA\ with respect to the nebular 
continuum. Part of this research was supported by PICS (Programme 
International de Cooperation Scientifique) \#247 of CNRS between France and 
Armenia. The financial support of International Association for the Promotion 
of Cooperation with Scientists from the Independant States (INTAS) is 
gratefully acknowledged. A.R. Petrosian would like to thank the warm 
hospitality of Institut d'Astrophysique de Paris and Observatoire de Marseille
during his stay. We also thank Jean Mouette for his help during the
preparation of the figures.}

\end{document}